\documentclass[pre,aps,amssymb,amsmath,superscriptaddress,twocolumn]{revtex4}

\usepackage{graphicx}% Include figure files
\usepackage{dcolumn}% Align table columns on decimal point
\usepackage{bm}% bold math
\usepackage{epstopdf}
\newcount\driver
\newcount\bozza

\font\cs=cmcsc10 scaled\magstep1
\font\ottorm=cmr8 scaled\magstep1 \font\msxtw=msbm10
scaled\magstep1                  \font\euftw=eufm10
scaled\magstep1 \font\msytw=msbm10 scaled\magstep1
\font\msytww=msbm8 scaled\magstep1 \font\msytwww=msbm7
scaled\magstep1                 \font\indbf=cmbx10 scaled\magstep2
\font\grbold=cmmib10 scaled\magstep1
\font\amit=cmmi7 \def\sf{\textfont1=\amit} \font\bigtenrm=cmr10
scaled \magstep2               \font\bigteni=cmmi10 scaled
{\count255=\time\divide\count255 by 60
\xdef\hourmin{\number\count255}
        \multiply\count255 by-60\advance\count255 by\time
   \xdef\hourmin{\hourmin:\ifnum\count255<10 0\fi\the\count255}}

\let\a=\alpha \let\b=\beta    \let\g=\gamma     \let\d=\delta     \let\e=\varepsilon
\let\z=\zeta  \let\h=\eta     \let\th=\vartheta \let\k=\kappa     \let\l=\lambda
\let\m=\mu    \let\n=\nu      \let\x=\xi        \let\p=\pi        \let\r=\rho
\let\s=\sigma \let\t=\tau     \let\f=\varphi    \let\ph=\varphi   \let\c=\chi
\let\ps=\psi  \let\y=\upsilon \let\o=\omega     \let\si=\varsigma
\let\G=\Gamma \let\D=\Delta   \let\Th=\Theta    \let\L=\Lambda    \let\X=\Xi
\let\P=\Pi    \let\Si=\Sigma  \let\F=\Phi       \let\Ps=\Psi
\let\O=\Omega \let\Y=\Upsilon

\def\PP{{\cal P}}\def\EE{{\cal E}}\def\MM{{\cal M}}\def\VV{{\cal V}}
\def\FF{{\cal F}}\def\HH{{\cal H}}\def\WW{{\cal W}}
\def\TT{{\cal T}}\def\NN{{\cal N}}\def\BB{{\cal B}}\def\ZZ{{\cal Z}}
\def\RR{{\cal R}}\def\LL{{\cal L}}\def\JJ{{\cal J}}\def\QQ{{\cal Q}}
\def\DD{{\cal D}}\def\AA{{\cal A}}\def\GG{{\cal G}}\def\SS{{\cal S}}
\def\OO{{\cal O}}\def\XXX{{\bf X}}\def\YYY{{\bf Y}}\def\WWW{{\bf W}}
\def\KK{{\cal K}}

\def\pp{{\bf p}}\def\qq{{\bf q}}\def\ii{{\bf i}}\def\xx{{\bf x}}
\def\aaa{{\bf a}} \def\bb{{\bf b}} \def\dd{{\bf d}}
\def\yy{{\bf y}}\def\kk{{\bf k}}\def\mm{{\bf m}}\def\nn{{\bf n}}
\def\zz{{\bf z}}\def\uu{{\bf u}}\def\vv{{\bf v}}\def\ww{{\bf w}}
\def\xxi{\hbox{\grbold \char24}} \def\bP{{\bf P}}\def\rr{{\bf r}}
\def\tt{{\bf t}}\def\bT{{\bf T}}

\def\ss{{\underline \sigma}}       \def\oo{{\underline \omega}}
\def\ee{{\underline \varepsilon}}  \def\aa{{\underline \alpha}}
\def\un{{\underline \nu}}          \def\ul{{\underline \lambda}}
\def\um{{\underline \mu}}          \def\ux{{\underline\xx}}
\def\uk{{\underline \kk}}          \def\uq{{\underline\qq}}
\def\uaa{{\underline \aaa}} \def\ub{{\underline\bb}}
\def\uc{{\underlinec}} \def\ud{{\underline\dd}}
\def\up{{\underline\pp}}           \def\ua{{\underline \a}}
\def\ut{{\underline t}}            \def\uxi{{\underline \xi}}
\def\umu{{\underline \m}}          \def\uv{{\underline\vv}}
\def\ue{{\underline \e}}           \def\uy{{\underline\yy}}
\def\uz{{\underline \zz}}
\def\uw{{\underline \ww}}          \def\uo{{\underline \o}}
\def\us{{\underline \s}}           \def\xxx{{\underline \xx}}
\def\kkk{{\underline\kk}}          \def\uuu{{\underline\uu}}
\def\udpr{{\underline\Dpr}}

\def\uu{\bf u}
\def\III{\hbox{\msytw I}}
\def\MMM{\hbox{\euftw M}}          \def\BBB{\hbox{\euftw B}}
\def\RRR{\hbox{\msytw R}}          \def\rrrr{\hbox{\msytww R}}
\def\rrr{\hbox{\msytwww R}}        

\def\NNN{\hbox{\msytw N}}          \def\nnnn{\hbox{\msytww N}}
\def\nnn{\hbox{\msytwww N}}        \def\ZZZ{\hbox{\msytw Z}}
\def\zzzz{\hbox{\msytww Z}}        \def\zzz{\hbox{\msytwww Z}}
\def\TTT{\hbox{\msytw T}}          \def\tttt{\hbox{\msytww T}}
\def\ttt{\hbox{\msytwww T}}        \def\EE{\hbox{\msytw E}}
\def\eeee{\hbox{\msytww E}}        \def\eee{\hbox{\msytwww E}}

\let\dpr=\partial
\let\circa=\cong
\let\bs=\backslash
\let\==\equiv
\let\txt=\textstyle
\let\io=\infty
\let\0=\noindent

\def\pagina{{\vfill\eject}}
\def\*{{\hfill\break\null\hfill\break}}
\def\bra#1{{\langle#1|}}
\def\ket#1{{|#1\rangle}}
\def\media#1{{\langle#1\rangle}}
\def\ie{\hbox{\it i.e.\ }}
\def\eg{\hbox{\it e.g.\ }}

\def\tilde#1{{\widetilde #1}}

\def\Dpr{\V\dpr\,}
\def\aps{{\it a posteriori}}
\def\lft{\left}
\def\rgt{\right}
\def\der{\hbox{\rm d}}
\def\la{{\langle}}
\def\ra{{\rangle}}
\def\norm#1{{\left|\hskip-.05em\left|#1\right|\hskip-.05em\right|}}
\def\tgl#1{\!\!\not\!#1\hskip1pt}
\def\tende#1{\,\vtop{\ialign{##\crcr\rightarrowfill\crcr
             \noalign{\kern-1pt\nointerlineskip}
             \hskip3.pt${\scriptstyle #1}$\hskip3.pt\crcr}}\,}
\def\otto{\,{\kern-1.truept\leftarrow\kern-5.truept\to\kern-1.truept}\,}
\def\fra#1#2{{#1\over#2}}

\def\sde{{\cs SDe}}
\def\wti{{\cs WTi}}
\def\osa{{\cs OSa}}
\def\ce{{\cs CE}}
\def\rg{{\cs RG}}

\def\lp{{\hskip-1pt:\hskip 0pt}}
\def\rp{{\hskip-1pt :\hskip1pt}}
\def\defi{{\buildrel \;def\; \over =}}
\def\apt{{\;\buildrel apt \over =}\;}
\def\nequiv{\not\equiv}
\def\Tr{\rm Tr}
\def\diam{{\rm diam}}
\def\sgn{\rm sgn}
\def\wt#1{\widetilde{#1}}
\def\wh#1{\widehat{#1}}
\def\hat#1{\wh{#1}}
\def\sqt[#1]#2{\root #1\of {#2}}

\def\ha{{\widehat \a}}\def\hx{{\widehat \x}}\def\hb{{\widehat \b}}
\def\hr{{\widehat \r}}\def\hw{{\widehat w}}\def\hv{{\widehat v}}
\def\hf{{\widehat \f}}\def\hW{{\widehat W}}\def\hH{{\widehat H}}
\def\hB{{\widehat B}}
\def\hK{{\widehat K}} \def\hW{{\widehat W}}\def\hU{{\widehat U}}
\def\hp{{\widehat \ps}}  \def\hF{{\widehat F}}
\def\bp{{\bar \ps}}
\def\hh{{\hat \h}}
\def\jm{{\jmath}}
\def\hJ{{\widehat \jmath}}
\def\hJ{{\widehat J}}
\def\hg{{\widehat g}}
\def\tg{{\tilde g}}
\def\hQ{{\widehat Q}}
\def\hC{{\widehat C}}
\def\hA{{\widehat A}}
\def\hD{{\widehat \D}}
\def\hDD{{\hat \D}}
\def\bl{{\bar \l}}
\def\hG{{\widehat G}}
\def\hS{{\widehat S}}
\def\hR{{\widehat R}}
\def\hM{{\widehat M}}
\def\hN{{\widehat N}}
\def\hn{{\widehat \n}}

\def\PP{{\cal P}}\def\EE{{\cal E}}\def\MM{{\cal M}}\def\VV{{\cal V}}
\def\FF{{\cal F}}\def\HH{{\cal H}}\def\WW{{\cal W}}
\def\TT{{\cal T}}\def\NN{{\cal N}}\def\BB{{\cal B}}\def\ZZ{{\cal Z}}
\def\RR{{\cal R}}\def\LL{{\cal L}}\def\JJ{{\cal J}}\def\QQ{{\cal Q}}
\def\DD{{\cal D}}\def\AA{{\cal A}}\def\GG{{\cal G}}\def\SS{{\cal S}}
\def\OO{{\cal O}}\def\AAA{{\cal A}}

\def\T#1{{#1_{\kern-3pt\lower7pt\hbox{$\widetilde{}$}}\kern3pt}}
\def\VVV#1{{\underline #1}_{\kern-3pt
\lower7pt\hbox{$\widetilde{}$}}\kern3pt\,}
\def\W#1{#1_{\kern-3pt\lower7.5pt\hbox{$\widetilde{}$}}\kern2pt\,}
\def\Re{{\rm Re}\,}\def\Im{{\rm Im}\,}
\def\lis{\overline}\def\tto{\Rightarrow}
\def\etc{{\it etc}} \def\acapo{\hfill\break}
\def\per{{\rm per}\,} \def\sign{{\rm sign}\,}
\def\indica{\leaders \hbox to 0.5cm{\hss.\hss}\hfill}
\def\guida{\leaders\hbox to 1em{\hss.\hss}\hfill}
\mathchardef\oo= "0521

\def\V#1{{\bf #1}}
\def\pp{{\bf p}}\def\qq{{\bf q}}\def\ii{{\bf i}}\def\xx{{\bf x}}
\def\yy{{\bf y}}\def\kk{{\bf k}}\def\mm{{\bf m}}\def\nn{{\bf n}}
\def\dd{{\bf d}}\def\zz{{\bf z}}\def\uu{{\bf u}}\def\vv{{\bf v}}
\def\xxi{\hbox{\grbold \char24}} \def\bP{{\bf P}}\def\rr{{\bf r}}
\def\tt{{\bf t}} \def\bz{{\bf 0}}
\def\ss{{\underline \sigma}}\def\oo{{\underline \omega}}
\def\xxx{{\underline\xx}}
\let\ciao=\bye
\def\qed{\raise1pt\hbox{\vrule height5pt width5pt depth0pt}}
\def\barf#1{{\tilde \f_{#1}}} \def\tg#1{{\tilde g_{#1}}}
\def\bq{{\bar q}} \def\bh{{\bar h}} \def\bp{{\bar p}} \def\bpp{{\bar \pp}}
\def\Val{{\rm Val}}
\def\indic{\hbox{\raise-2pt \hbox{\indbf 1}}}
\def\bk#1#2{\bar\kk_{#1#2}}
\def\tdh{{\tilde h}}

\def\RRR{\hbox{\msytw R}} \def\rrrr{\hbox{\msytww R}}
\def\rrr{\hbox{\msytwww R}} 
\def\NNN{\hbox{\msytw N}} \def\nnnn{\hbox{\msytww N}}
\def\nnn{\hbox{\msytwww N}} \def\ZZZ{\hbox{\msytw Z}}
\def\zzzz{\hbox{\msytww Z}} \def\zzz{\hbox{\msytwww Z}}
\def\TTT{\hbox{\msytw T}} \def\tttt{\hbox{\msytww T}}
\def\ttt{\hbox{\msytwww T}}

\def\ins#1#2#3{\vbox to0pt{\kern-#2 \hbox{\kern#1 #3}\vss}\nointerlineskip}

\newdimen\xshift \newdimen\xwidth \newdimen\yshift
\def\insertplot#1#2#3#4#5#6{%
\xwidth=#1pt \xshift=\hsize \advance\xshift by-\xwidth \divide\xshift by 2%
\begin{figure}[ht]
\vspace{#2pt} \hspace{\xshift}
\begin{minipage}{#1pt}
#3 \ifnum\driver=1 \griglia=#6
\ifnum\griglia=1 \openout13=griglia.ps \write13{gsave .2
setlinewidth} \write13{0 10 #1 {dup 0 moveto #2 lineto } for}
\write13{0 10 #2 {dup 0 exch moveto #1 exch lineto } for}
\write13{stroke} \write13{.5 setlinewidth} \write13{0 50 #1 {dup 0
moveto #2 lineto } for} \write13{0 50 #2 {dup 0 exch moveto #1
exch lineto } for} \write13{stroke grestore} \closeout13
\includegraphics{griglia.ps} \fi
\includegraphics{#4.ps}\fi%
\ifnum\driver=2 \fi
\end{minipage}
\caption{#5}
\end{figure}
}
\def\gtopl{\hbox{\msxtw \char63}}
\def\ltopg{\hbox{\msxtw \char55}}

\newdimen\shift \shift=-1.5truecm
\def\lb#1{%
\ifnum\bozza=1
\label{#1}\rlap{\hbox{\hskip\shift$\scriptstyle#1$}}
\else\label{#1} \fi}

\def\be{\begin{equation}}
\def\ee{\end{equation}}
\def\bea{\begin{eqnarray}}\def\eea{\end{eqnarray}}
\def\bean{\begin{eqnarray*}}\def\eean{\end{eqnarray*}}
\def\bfr{\begin{flushright}}\def\efr{\end{flushright}}
\def\bc{\begin{center}}\def\ec{\end{center}}
\def\bal{\begin{align}}\def\eal{\end{align}}
\def\ba#1{\begin{array}{#1}} \def\ea{\end{array}}
\def\bd{\begin{description}}\def\ed{\end{description}}
\def\bv{\begin{verbatim}}\def\ev{\end{verbatim}}
\def\nn{\nonumber}
\def\Halmos{\hfill\vrule height10pt width4pt depth2pt \par\hbox to \hsize{}}
\def\pref#1{(\ref{#1})}
\def\Dim{{\bf Dim. -\ \ }} \def\Sol{{\bf Soluzione -\ \ }}
\def\virg{\quad,\quad}
\def\bsl{$\backslash$}

\def\ins#1#2#3{\vbox to0pt{\kern-#2 \hbox{\kern#1 #3}\vss}\nointerlineskip}

\newdimen\xshift \newdimen\xwidth \newdimen\yshift
\newcount\griglia

\def\insertplot#1#2#3#4#5#6{%
\xwidth=#1pt \xshift=\hsize \advance\xshift by-\xwidth \divide\xshift by 2%
\begin{figure}[ht]
\vspace{#2pt} \hspace{\xshift}
\begin{minipage}{#1pt}
#3 \ifnum\driver=1 \griglia=#6
\ifnum\griglia=1 \openout13=griglia.ps \write13{gsave .2
setlinewidth} \write13{0 10 #1 {dup 0 moveto #2 lineto } for}
\write13{0 10 #2 {dup 0 exch moveto #1 exch lineto } for}
\write13{stroke} \write13{.5 setlinewidth} \write13{0 50 #1 {dup 0
moveto #2 lineto } for} \write13{0 50 #2 {dup 0 exch moveto #1
exch lineto } for} \write13{stroke grestore} \closeout13
\includegraphics{griglia.ps} \fi
\includegraphics{#4.ps}\fi%
\ifnum\driver=2 \fi
\end{minipage}
\caption{#5}
\end{figure}
}
\def\gtopl{\hbox{\msxtw \char63}}
\def\ltopg{\hbox{\msxtw \char55}}

\newdimen\shift \shift=-1.5truecm
\def\lb#1{%
\label{#1}\rlap{\hbox{\hskip\shift$\scriptstyle#1$}}
\else\label{#1} \fi}

\def\be{\begin{equation}}
\def\ee{\end{equation}}
\def\bea{\begin{eqnarray}}\def\eea{\end{eqnarray}}
\def\bean{\begin{eqnarray*}}\def\eean{\end{eqnarray*}}
\def\bfr{\begin{flushright}}\def\efr{\end{flushright}}
\def\bc{\begin{center}}\def\ec{\end{center}}
\def\bal{\begin{align}}\def\eal{\end{align}}
\def\ba#1{\begin{array}{#1}} \def\ea{\end{array}}
\def\bd{\begin{description}}\def\ed{\end{description}}
\def\bv{\begin{verbatim}}\def\ev{\end{verbatim}}
\def\nn{\nonumber}
\def\Halmos{\hfill\vrule height10pt width4pt depth2pt \par\hbox to \hsize{}}
\def\pref#1{(\ref{#1})}
\def\Dim{{\bf Dim. -\ \ }} \def\Sol{{\bf Soluzione -\ \ }}
\def\virg{\quad,\quad}
\def\bsl{$\backslash$}

\usepackage{amsmath}
\usepackage{amsfonts}
\usepackage{amssymb}
\usepackage{epstopdf}

\font\msytw=msbm9 scaled\magstep1 \font\msytww=msbm7
scaled\magstep1 \font\msytwww=msbm5 scaled\magstep1
\font\cs=cmcsc10

\let\a=\alpha \let\b=\beta  \let\g=\gamma  \let\d=\delta
\let\e=\varepsilon
\let\z=\zeta  \let\h=\eta   \let\th=\theta \let\k=\kappa \let\l=\lambda
\let\m=\mu    \let\n=\nu    \let\x=\xi     \let\p=\pi    \let\r=\rho
\let\s=\sigma \let\t=\tau   \let\f=\varphi \let\ph=\varphi\let\c=\chi
\let\ps=\Psi  \let\y=\upsilon \let\o=\omega\let\si=\varsigma
\let\G=\Gamma \let\D=\Delta  \let\Th=\Theta\let\L=\Lambda \let\X=\Xi
\let\P=\Pi    \let\Si=\Sigma \let\F=\Phi    \let\Ps=\Psi
\let\O=\Omega \let\Y=\Upsilon

\def\PPP{{\cal P}}\def\EE{{\cal E}}\def\MM{{\cal M}} \def\VV{{\cal V}}
\def\FF{{\cal F}} \def\HHH{{\cal H}}\def\WW{{\cal W}}
\def\TT{{\cal T}}\def\NN{{\cal N}} \def\BBB{{\cal B}}\def\III{{\cal I}}
\def\RR{{\cal R}}\def\LL{{\cal L}} \def\JJ{{\cal J}} \def\OO{{\cal O}}
\def\DD{{\cal D}}\def\AAA{{\cal A}}\def\GG{{\cal G}} \def\SS{{\cal S}}
\def\KK{{\cal K}}\def\UU{{\cal U}} \def\QQ{{\cal Q}} \def\XXX{{\cal X}}

\def\qq{{\bf q}} \def\pp{{\bf p}}
\def\vv{{\bf v}} \def\xx{{\bf x}} \def\yy{{\bf y}} \def\zz{{\bf z}}
\def\aa{{\bf a}}\def\hh{{\bf h}}\def\kk{{\bf k}}
\def\mm{{\bf m}}\def\PP{{\bf P}}

\def\dd{{\boldsymbol{\delta}}}

\def\ddd{\boldsymbol{\d}}
\def\TTTT{\mathbf{T}}

\def\nn{\nonumber}
\def\us{\underset}
\def\os{\overset}

\def\RRR{\hbox{\msytw R}} \def\rrrr{\hbox{\msytww R}}
\def\rrr{\hbox{\msytwww R}}
\def\NNN{\hbox{\msytw N}} \def\nnnn{\hbox{\msytww N}}
\def\nnn{\hbox{\msytwww N}} \def\ZZZ{\hbox{\msytw Z}}
\def\zzzz{\hbox{\msytww Z}} \def\zzz{\hbox{\msytwww Z}}
\def\TTT{\hbox{\msytw T}}

\def\\{\hfill\break}
\def\={:=}
\let\io=\infty
\let\0=\noindent\def\pagina{{\vfill\eject}}
\def\media#1{{\langle#1\rangle}}
\let\dpr=\partial
\def\sign{{\rm sign}}
\def\const{{\rm const}}
\def\tende#1{\,\vtop{\ialign{##\crcr\rightarrowfill\crcr\noalign{\kern-1pt
    \nointerlineskip} \hskip3.pt${\scriptstyle #1}$\hskip3.pt\crcr}}\,}
\def\otto{\,{\kern-1.truept\leftarrow\kern-5.truept\to\kern-1.truept}\,}
\def\defin{{\buildrel def\over=}}
\def\wt{\widetilde}
\def\wh{\widehat}
\def\to{\rightarrow}
\def\la{\left\langle}
\def\ra{\right\rangle}
\def\qed{\hfill\raise1pt\hbox{\vrule height5pt width5pt depth0pt}}
\def\Val{{\rm Val}}
\def\ul#1{{\underline#1}}
\def\lis{\overline}
\def\V#1{{\bf#1}}
\def\be{\begin{equation}}
\def\ee{\end{equation}}
\def\bp{\begin{pmatrix}}
\def\ep{\end{pmatrix}}
\def\bea{\begin{eqnarray}}
\def\eea{\end{eqnarray}}
\def\nn{\nonumber}
\def\pref#1{(\ref{#1})}
\def\ie{{\it i.e.}}
\def\lb{\label}
\def\eg{{\it e.g.}}

\def\Tr{\mathrm{Tr}}
\def\eu{\mathrm{e}}

\newtheorem{lemma}{Lemma}[section]
\newtheorem{theorem}{Theorem}[section]

\begin{document}

\title{Coupled identical localized fermionic chains with quasi-random disorder}

\author{Vieri Mastropietro}

\address{
Universit\'a di Milano, Via C. Saldini 50, 20133, Milano, Italy
}

\begin{abstract} 
We analyze the ground state localization properties of 
an array of identical interacting spinless fermionic chains with quasi-random disorder, 
using non-perturbative Renormalization Group methods. In the single or
two chains case localization persists while 
for a larger number of chains  
a different qualitative behavior is generically expected, unless 
the many body interaction is vanishing.
This 
is due to number theoretical properties of the frequency, similar to the ones assumed in KAM theory, and 
cancellations due to Pauli principle which in the single or two chains case imply that all the effective interactions are 
irrelevant; in contrast for a larger number of chains relevant effective interactions are present.
\end{abstract}


\maketitle

\renewcommand{\thesection}{\arabic{section}}

\renewcommand{\thesection}{\arabic{section}}

\section{Introduction}

A quantum system in which 
disorder-induced localization 
\cite{A} persists in presence
of a interaction
is said to be in a 
Many Body Localized (MBL) phase. 
While normal systems are expected to approach asymptotically a thermal state (due to interaction "the system acts as his own bath"),
this does not happen in a MBL phase \cite{NH}, \cite{NH1},\cite{HL}, a fact with
deep 
theoretical and technological implications. However
the interplay of disorder and interaction produces a complex behavior \cite{FA},  \cite{GGG} and
the existence itself of 
a MBL phase is quite a non trivial property which is under deep investigation. 

In the case of random disorder MBL was   
established order by order by formal series in any dimension \cite{Ba}, \cite{Ba1},\cite{Ba2}, but this does not exclude 
delocalization due to the possible divergence of the expansions.
In one dimension a non-perturbative proof of MBL has been reached \cite{loc4}, \cite{loc4a}, but it relies on 
a still unproven assumption. Numerical evidence of MBL in one dimensional lattices has been obtained in  \cite{PH1},\cite{PH4}, \cite{z1}. 

Also quasi-random disorder in one dimension produces localization in the single particle case, as 
found in \cite{AA} and rigorously proved in 
\cite{FS}, \cite{Ia}. In presence of interaction, a non perturbative proof of
ground state localization has been achieved in 
\cite{M}.  Numerical
evidence of MBL with quasi random disorder has been found
in \cite{H4} ,\cite{I}, \cite{B3}, \cite{R}, \cite{PP}. One dimensional systems of particles with quasi-random disorder can be realized in cold-atoms experiments 
\cite{B} and evidence of MBL was claimed. 

As a natural step toward higher dimensions we consider an array of interacting fermionic chains with Aubry-Andre' quasi random disorder \cite{AA}
and coupled by an hopping term.
Such model (with spinful fermions) has been realized in cold atoms experiments  in  \cite{B1}. 
We call $x=0,\pm 1,\pm 2,..$ the coordinates of the infinite chain and $y=0,..,L$ the coordinates labeling the chains, and we consider a system of $N$ spinless fermions with
Hamiltonian 
\be
H_N=
\sum_{i=1}^N H_{A}(x_i)+J_\perp 
\sum_{i=1}^N \nabla_{y_i}
+U\sum_{i,j\in 1}^N v(x_i-x_j)\label{11}
\ee
where $v(x_i-x_j)=\d_{y,x+1}$ and
$H_{A}$ is the Aubry-Andre' Hamiltonian 
\be
H_{A}(x)=
J \nabla_{x}+\D\cos (2\pi(\o x+\th))\label{12}
\ee
and $\nabla_{z} f(z)= f(z-1)+f(z+1)-2 f(z)$; periodic boundary conditions are imposed in $y$. The Hamiltonian 
\pref{11} describes $L$ fermionic chains, with identical disorder, intra-chain hopping $J$
, intra-chain interaction $U$ and inter-chain hopping $J_\perp$.
If $J_\perp=U=0$ the system reduces to several uncoupled Aubry-Andre' models \cite{AA}. 
The behavior of the eigenfunctions of $H_A$ 
\pref{12}
depends crucially on the ratio ${\D\over J}$ between the disorder and the hopping;
if ${\D\over J}<2$ the eigenfunctions are quasi-Bloch extended waves while for 
${\D\over J}>2$ are exponentially decaying 
and Anderson localization occurs \cite{FS}, \cite{Ia}.  A metal-insulator transition is therefore present varying the strength of the disorder, a feature making quasi-random disorder somewhat similar to random disorder in three dimensions. 

The question we address in this paper is if a localized phase persists
in the array described by \pref{11}, and how the behavior depends on
the interplay between the hopping $J_\perp$, the interaction $U$ and the number of chains $L$. 
The main theoretical difficulty
is that localization is a {\it non-perturbative}
phenomenon; the presence or absence of localization is related
to the convergence or divergence of the series, driven by  {\it small divisors} which can produce dangerous factorials. Information is carried by high orders and instability is not signaled by divergences at low orders, as it happens in quantum field theory. 

In the single particle case $H_A$
the small divisors are similar to the ones in the series in Kolmogorov-Arnold-Moser (KAM) theory, whose convergence
implies stability in close-to-integrable system, while divergence is related to the onset of chaos. Indeed the eigenfunctions of \pref{12}, 
can be written in series of $J$ and divisors are of the form
$\phi_x-\phi_y$ with $x\not=y$, with
$\phi_x=\D\cos (2\pi(\o x+\th))$. In order to get convergence, and as a consequence localization, one needs to assume number theoretical conditions, called Diophantine (see below), to control the size of $||(\o n)||$
and  $||(\o n+2\th)||$, with $||.||$ the norm on the side $1$ torus, see 
\cite{FS}, \cite{Ia}. Such Diophantine conditions are the same assumed in KAM theory.
In presence of interaction the small divisors in the
expansion for the $N$-particle eigenfunctions are much more complex; they are of the form $E_N(\underline x)-E_N(\underline y)$, with 
$E_N(\underline x)=\sum_{i=1}^N \phi_{x_i}$. No number theoretical
condition is known to control them for $N>1$ \cite{E} (for  
the $N=2$ case see \cite{S}).

Even if the construction of all the eigenfunctions of \pref{11} for a generic $N$ is outside the present analytical possibilities, we can 
analyze the problem using a different approach, introduced in \cite{M}:
we do not consider the expansion for the eigenfuctions but
we compute in the thermodynamical limit $N\to\io$
the grand canonical correlations, which at zero temperature becomes the ground state correlations. 
This approach allows to take advantage of fermionic cancellations 
and non-perturbative and rigorous information on localization
of systems with an infinite number of particles, even if limited to the ground state, can be obtained.
The correlations are written 
as Grassmann integrals
which are analyzed via exact fermionic Renormalization Group (RG) methods;
one integrates out the degrees of freedom with smaller and smaller energy 
obtaining a sequence of effective interactions, sum of terms which are all
{\it relevant} in the RG sense, independently from the number of fields. The presence of an infinite number of relevant processes seems to say that an RG approach is hopeless; however
by exploiting number theoretical properties of the frequency of the incommensurate disorder
it is possible to show that a huge class of effective interactions, called {\it non resonant}, are indeed {\it irrelevant}.
In contrast with the single chain problem, in which Diophantine conditions are sufficient, 
here one needs also other conditions called in KAM theory the first and second {\it Melnikov conditions}.
While in absence of interaction the structure of Feynman graphs is rather simple, the presence of interaction $U\not=0$ complicates considerably the problem; one has a combination of small divisors and loops, which are 
absent in non interacting or KAM-like problems. Other dangerous factorials,
in addition to the ones produced by small divisors, are produced by combinatorics related to the number of graphs; they are controlled by cancellations due to the fermionic sign cancellations. 

A renormalized expansion is obtained in terms of the running coupling constants corresponding to the resonant terms.
As usual in RG, the physical properties depend on their flow; if the running coupling constants do not exit from the convergence radius the interacting theory is analytically close to the free one, so that localization persists in presence
of interaction. The 
flow dramatically depends on the number of chains. 
In the two chain problem there are no relevant effective quartic interactions, the only relevant couplings being
quadratic, as in the single chain problem; localization
persists in the ground state in presence of interaction. 

On the contrary, with an higher number of chains the quartic 
terms are relevant, and their size increase iterating the RG; therefore 
a different qualitative behavior is generically expected, unless 
the many body interaction is vanishing, where localization still persists.
%The results appear then in good qualitative agreement with the experiments in \cite{B1} in whih a large number of chain %is considered 
%; in addition, we predict that in the two chain case localization persists in the spinless case.  

The content of this paper is the following.
In \S 2 we present the main results. In \S 3 we perform an exact RG analysis and we show the irrelevance of the non resonant terms. In \S 4 we identify the relevant and marginal 
terms and study the corresponding flow, and 
in \S 5 we get our main results discussing the convergence of the expansion.
Finally in \S 6 the main conclusions are presented.

\section{Main result}

We consider the grand canonical averages $<O>
=\sum_N {\Tr_N e^{-\b (H_N-\m N)} O\over Z}$, with 
$Z=\sum_N \Tr_N e^{-\b (H_N-\m N)}$;  the thermodynamic limit is taken sending the chain length to infinity keeping the number of chains $L$ finite. The Fock space Hamiltonian is
\bea 
&&H=J\sum_{x,y} (a^+_{x+ 1,y} a^-_{x,y}+ a ^+_{x-1,y} a^-_{x,y}) +\nn\\
&&\D\sum_{x,y} 
\cos (2\pi \o x) a^+_{x,y} a^-_{x,y}+U\sum_{x,y}
a^+_{x,y} a^-_{x,y} a^+_{x+1,y}
a^-_{x+1,y}\nn\\
&&+J_\perp  \sum_{x,y} (a^+_{x,y+1} a^-_{x,y}+ a ^+_{x,y} a^-_{x,y+1}) 
\label{1.1}\eea
and we assume for definiteness the phase of the disorder equal to zero.
%It is well known that the interaction produces a renormalization of the %chemical potential; that is, if we fix the density, the corresponding value of %the chemical potential depends from the interaction. We will then write $%\m=\m_0+\n$, where $\n$ is a counterterm to be suitable chosen so that %the chemical potential of the interacting theory is $\m_0$. We call  
%$\m_0=\D\cos 2\pi \o \bar x$.
It is convenient to write
$a^\pm_{x,y}=
{1\over L}\sum_{l}e^{\pm i l y} \hat a^\pm_{x,l}$, where $l=2\pi{n\over L}$
with $n=0,..,L-1$ so that the Hamiltonian can be rewritten in the following way
 %r
\bea 
&&H=J{1\over L}\sum_{x,l} (\hat a^+_{x+ 1,l} \hat a^-_{x,l}+ \hat a^+_{x-1,l} \hat a^-_{x,l}) +\\
&&{\D\over L}\sum_{x,l} 
\cos (2\pi \o x) \hat a^+_{x,l} \hat a^-_{x,l}+{J_\perp\over L} \sum_{x,l} \cos l (\hat a^+_{x,l} \hat a^-_{x,l}+ 
\hat a^+_{x,l} \hat a^-_{x,l})\nn\\
&&U\sum_{x}{1\over L^4}\sum_{l_1,l_2,l_3,l_4}\hat a^+_{x,l_1} \hat a^-_{x,l_2} \hat a^+_{x+1,l_3}
\hat a^-_{x+1,l_4}\d(l_1-l_2+l_3-l_4)
\label{1.1a}\nn\eea
We focus on the 2-point function $< \hat a^-_{\xx,l}  \hat a^+_{\zz, l}>$, where  
$< O>=
{\Tr e^{-\b(H-\m N)} T O \over \Tr e^{-\b(H-\m N)} }$, $T$ is the time ordering and $\hat a^\pm_{\xx,l}=e^{(H-\m N)x_0} \hat a^\pm_{x,l} e^{-(H-\m N)x_0}$ and 
$\xx=(x_0,x)$. 
In the {\it molecular limit} $U=J=0$ one has (setting $\D=1$ for definiteness)
\be
H_0-\m N={1\over L}\sum_{x,l} 
(\cos (2\pi \o x)-\m_l)  \hat a^+_{x,l} \hat a^-_{x,l}
\ee
with 
\be
\m_l=\m+J_\perp \cos l\equiv\cos 2\pi \bar x_l
\ee
so that, calling $\m=\cos(2\pi\o \bar x)$, than $\bar x_l=\bar x+a \cos l J_\perp$ with $a^{-1} =\sin 2\pi \bar x+O(J_\perp)$. In this limit the system
is uncoupled with an $l$-dependent chemical potential for any chain. The ground state occupation number is $=1$
for $\cos (2\pi \o x)<\m_l$ and $0$ for $\cos (2\pi \o x)>\m_l$.
The 2-point function
$< \hat a^-_{\xx,l} \hat a^+_{\yy,l}>|_{U=J=0} \equiv g_l(\xx,\yy)$
is equal to
\be g_l(\xx,\yy)=
\d_{x,y}{1\over\b}\sum_{k_0={2\pi\over\b}(n_0+{1\over 2})}
\hat g_l(x,k_0)e^{-i k_0 (x_0-y_0)}\label{prop} \ee
with \bea&&\hat g_l(x, k_0)=
\int_0^\b d\t e^{i \t k_0}{e^{-\t (\cos 2\pi \o x-\m_l)} \over 1+e^{-\b (\cos 2\pi \o x-\m_l))
}}=\nn\\
&&{1\over -i k_0+\cos 2\pi \o x-\cos 2\pi \o \bar x_l}\label{bb1}\eea
The 2-point function is perfectly localized in the chain direction (the 2-point function is vanishing if $x\not=y$), but not on the transversal direction in the coordinate space. Assume that $\bar x_l$ is not a point of the lattice, so that the propagator
\pref{bb1} is never singular.
As $\o$ is an irrational number, $\o x$ modulo $1$ fills densely the set $[-1/2,1/2)$, and in particular it can be arbitrarily close to $\pm \o\bar x_l$. If we set $x=x' +\r_l \bar x_l$ then for $(\o x')_{mod 1}$ small, $\r_l=\pm$ 
\be \hat g_l(x, k_0)\sim {1\over -i k_0+v_l \r_l (\o x')_{mod 1}}\label{bb}
\ee
and $v_l=\sin 2\pi \o \bar x_l$. The expansion of the 2-point function in terms of $J,U$ can be represented in terms of Feynman graphs, expressed by product of propagators $\hat g_l(x, k_0)$; on each line of the diagram is associated a coordinate
$x$ and the difference of lines coming in or out from the vertex $J$ is $\pm 1$, while from an $U$ vertex is $0,\pm 1$. 
Note the similarity of \pref{bb1} with the 2-point function in the free fermion limit $\D=U=J_\perp=0$, $J=1$ which in Fourier space 
is given by $1/-i k_0+\cos k-\m$. If
$k=k'\pm p_F$, $\m=\cos p_F$ the free fermion propagator is asymptotically given by 
$1/-i k_0\pm v_F k'$, which is the well known {\it Luttinger liquid} propagator.
 $p_F$ are called Fermi momenta and by analogy we can call $\pm  \bar x_l$ the  
{\it Fermi coordinates}. 

The expansion in $J,U$ around the molecular limit
is convergent at finite temperature, as the temperature acts as an infrared cut-off, and the main issue is to get the zero temperature limit.
We expect that the interaction produces a renormalization of the chemical potential, and it is convenient to fix the renormalized chemical potential to a $J, U$-independent value; this corresponds to fix the density of the interacting system. We therefore write
\be
\m_l=\cos 2\pi \o \bar x_l+\n_l
\ee
where $\n_l$ is a counterterm to be fixed so that the chemical potential of the interacting theory is $\cos 2\pi \o \bar x_l$. In order to understand
the behavior at high orders one needs to exploit some number theoretical property of $\o$; in particular, as in the analysis of the Aubry-Andre' model, we assume that
the frequency $\o$ is a Diophantine number, verifying the property
\be || \o x||\ge C_0
|x|^{-\t}\quad
\forall x\in  \mathbb{Z}/\{0\}
\label{d}\ee with $||.||$ is the norm on the one dimensional torus.  Such a property, saying roughly speaking that 
$\o$ is a ``good'' irrational, is not restrictive as 
Diophantine numbers have full measure. 
As an example, the golden ratio $\o={\sqrt{5}+1\over 2}$ verifies \pref{d} with $\t=1$ and $C_0={3+\sqrt{5}\over 2}$. The Diophantine condition will ensure that a process involving fermions living close to $(\o \bar x_l)$ involves fermions with a huge difference of coordinates.

In addition one has to assume a diophantine  condition on the chemical potential (equivalently one can assume a similar condition on $\th$), namely 
\be ||\o x\pm 2\o\bar x||
\ge C_0 |x|^{-\t}\quad
\forall x\in  \mathbb{Z}/\{0\}
\label{d1}\ee with $||.||$ is the norm on the one dimensional torus. Such condition 
says $\bar x$ is incommensurate with $\o$. In the decoupled case $J_\perp=0$ this implies that  $|\hat g_l(x, k_0)|\le C |x|^\t$.
Our main result is the following. 
\vskip.3cm
{\it If $U,J_\perp, J$ are small, $J_\perp\not=0$ belongs to a set of large relative measure and if $\o,\bar x$ verify
\pref{d} and \pref{d1}, for a suitable $\n_l$, then:

a)
if $L=2$ for $\b\to\io$ then for any integer $N$ and a suitable constant $C_N$
\be
|< \hat a^-_{\xx,l} \hat a^+_{\yy,l}>|
\le e^{-\x |x-y|} {C_N |\log \D| \over 1+(\D|x_0-y_0|)^N}\label{zz}
\ee
with 
$\x=|\log\e|$
and $\D=
(1+\min(|x|,|y|))^{-\t}$. 
\vskip.3cm
b)If $L\ge 3$, $U=0$ then for $\b\to\io$ \pref{zz} holds.
\vskip.3cm
c)If $L\ge 3$, then \pref{zz} holds
for $\b |U |\le 1$, with 
$\x=\max(|\log\e|,\b^{-1})$, $\D=\max(
(1+\min(|x|,|y|))^{-\t},\b^{-1})$}
\vskip.2cm
In the case of two chains (case a) the 2-point function decays at zero temperature exponentially in the direction of the chains, and a very weak decay is present in the imaginary time direction
(faster than any power but with rate decreasing increasing $x,y$); this is very similar to what happens in the single chain case and indicates localization of the ground state with or without interaction. In contrast, for a greater number of chains
the interaction produces a qualitative difference; 
in absence of many body interaction, zero temperature exponential decay 
is found for any number of chains (case b) while 
in presence of interaction convergence of the expansion holds only up to a finite temperature (case c). The reason is that when $L\ge 3$ there are extra relevant terms increasing iterating the RG, and this has the effect that convergence holds only for temperatures not too small; as usual, the presence of diverging directions in the RG flow is expected to signal an instability of the system. This 
provides an explanation of the behavior observed 
in cold atoms experiments \cite{B1}, in which absence of localization is found in an array of chains (except when there is no interaction, when localization is found), and localization in the single chain case; moreover, we find localization with two chains in the spinless case, a prediction in principle accessible to future experiments.

\section{Renormalization Group analysis}

The 2-point function is obtained by the second derivative of the generating function 
\be
e^{W(\phi)}=\int P(d\psi)e^{V(\psi)+(\psi,\phi)}\label{ww}
\ee
with
\bea
&&V={1\over L}\sum_l \int d\xx J (
\psi^+_{\xx,l} \psi^-_{\xx+{\bf e_1},l}+\psi^+_{\xx+{\bf e_1},l}\psi^-_{\xx,l})+\nn\\
&&
\int d\xx {U\over L^4}\sum_{\underline l}
\psi^+_{\xx,l_1}\psi^-_{\xx,l_2} \psi^+_{\xx+{\bf e_1},l_3}
\psi^-_{\xx+{\bf e_1},l_4}\d(l_1-l_2+l_3-l_4)\nn\\
&&+{1\over L}\sum_l \n_l \int d\xx 
\psi^+_{\xx,l} \psi^-_{\xx,l}
\eea
where $\psi$ are grassmann variables, $\phi$ is the external source, $\int d\xx=\int dx_0\sum_x$,
${\bf e}_1=(0,1)$ and $P(d\psi)$ is the fermionic integration with propagator 
\pref{prop}

We introduce
a cut-off smooth function $\chi_\r(k_0,x)$ which is non vanishing for
$\sqrt{k_0^2+(v_l(\o(x-\r \bar x_l)_{\rm mod. 1})^2)}\le \g$, where $\r=\pm 1$ and $\g>1$
is a suitable constant (to be fixed below); therefore we can write the propagator as 
\be
\hat g_l(\xx)=\hat g^{(u.v.)}_l(\kk)+\sum_{\r=\pm}\hat g_{\r,l}(\kk)
\ee where \be \hat g_{\r,l}(k_0, x)={\chi_\r(k_0,x)\over -i k_0+\cos(2\pi(\o x))
-\cos(2\pi(\o \bar x_l))}\ee 
and correspondingly
 $\psi_{k_0, x, l}=\psi^{(u.v.)}_{k_0, x, l}+\sum_{\r=\pm 1} \psi_{\r; k_0, x,l}$. This simply says that we are rewriting 
the fermionic field as sum of two independent fields living close to one of the Fermi points, up to a regular field.
We can further decompose \be \hat g_{\r,l}(k_0,
 x)=\sum_{h=h_\b}^0 \hat g^{(h)}_{\r,l}( k_0,x)\ee with $-h_\b\sim \log \b$,
$\hat g^{(h)}_{\r,l}(k_0,x)$ similar to $\hat g_{\r,l}(k_0,x)$ with $\chi$ replaced by $f_h$ 
with
$f_h(k_0,\o x')$ non vanishing in a region $\sqrt{k_0^2+(v_l (\o x')_{\rm mod 1})^2}\sim \g^h$ with $x=x'+\r\bar x_l$.

After the integration of the fields $\psi^{(u.v.)}, \psi^{(0)},..,\psi^{(h+1)}$ the generating function has the form
\be
e^{W(\phi)}=\int P(d\psi^{\le h})e^{V^{(h)}(\psi)+B^{(h)}(\psi,\phi)}\label{hh}
\ee
where $P(d\psi^{\le h})$ has propagator $g_{\r,l}^{(\le h)}=\sum_{k=-\io}^h g_{\r,l}^{(k)}$ and $V^{(h)}(\psi)$
is given by sum of terms
\bea
&&\sum_{x'_1}
\int dx_{0,1}...\int dx_{0,m}{1\over L^m}\sum_{l_1,..,l_m}
W_{m,\underline l}^{(h)}(x'_1, \underline x, )\nn\\
&&\d(\sum_i \e_i l_i)
\psi^{\e_1(\le h)}_{\r_1;x_{0,1},x'_1,l_1}... \psi^{\e_m(\le h)}_{\r_m;x_{0,m},x'_m,l_m}
\label{0,1}\eea
where the kronecker deltas in the propagators imply that a single sum over $x$ is present; the kernels $W_m^{(h)}$ are sum of Feynman diagrams obtained connecting vertices $J$, $U$
or $\n$ with propagators $g^{(k)}$ with $k>h$. Similarly $B^{(h)}$ is given by a similar expression with the only difference that some of the external lines are associated to $\phi$ fields. The scaling dimension
of the theory can be obtained by the bounds
\be\int dx_0 |g^{(h)}_\r(x_0, x)|\le C\g^{-h}\quad|g^{(h)}_\r(x_0, x)|\le C\label{jjj}\ee  
The persistence or not of localization is related to the presence or lack of convergence, that is the behavior at high orders; we need therefore to remind some basic tool of renormalization theory, which are  crucial to avoid the well known problem of "overlapping divergences". 
Given a Feynman graph, one
considers a maximally connected subset of lines corresponding 
to propagators 
with scale $h\ge h_v$ with at least a scale $h_v$, and we call it {\it cluster} $v$ (for more details, see \cite{M1}); 
the external lines have scale smaller then $h_v$. Therefore to each Feynman graph is associated a hierarchy of 
clusters; inside each cluster $v$ there are $S_v$ maximal clusters, that is clusters   
contained only in the cluster $v$ and not in any smaller one, or trivial clusters given by a single vertex. The clusters therefore identify the subdiagrams which one needs to renormalize, as the ones containing propagators living at energy scales greater than the ones outside them.

Each of such $S_v$ clusters are connected 
by a tree of propagators with scale $h_v$; by integrating the propagators over time and using 
\pref{jjj} we get 
that each graph of order $n$ contributing to $W^{(h)}_m$ is bounded at fixed scale by, if $\e=\max(|J|,|U|)$ 
\be C^n \e^n 
\prod_v \g^{-h_v(S_v-1)}\label{paz}\ee
where $v$ are the clusters (not end-points) and
$h_v\le 0$. From the above estimate we see that the scaling dimension of any contribution to the effective potential has the same positive scaling dimension (independently from the number of fields)
\be
D=1
\ee
In other words all the effective interactions are {\it relevant} in the RG sense and the theory is {\it non-renormalizable}; indeed to the effective potential graphs with all the assignments of scales contribute and 
from \pref{paz} the sum over scales gives an infinite result.
However, it turns out, as a consequence of number theoretical properties
of the quasi random disorder, that a huge class of terms are indeed {\it irrelevant}. In a large relative measure set of  
$J_\perp$ one has
\be ||\o x\pm 2 \o\bar x_l||
\ge C_0 |x|^{-\t'}\quad\quad
\forall x\in  \mathbb{Z}/\{0\}
\label{d33}\ee
and
\be ||\o x\pm \o\bar x_l\pm  \o\bar x_{l'} ||
\ge C_0 |x|^{-\t'}\quad\quad
\forall x\in  \mathbb{Z}/\{0\}
\label{d34}\ee
Conditions \pref{d33} and  \pref{d34} 
are known in KAM theory as the first and second {\it Melnikov conditions}.
The first condition is used to bound the propagator; using that $||\o x'||=||\o x-\r \o \bar x_l||=||\o 2  x-2 \r \o \bar x_l||$ for $||\o x'||$ small
then $|\hat g^{(h)}(k_0,x)|\le C |x|^\t$. 
The second condition is used to show the irrelevance
of a number of terms in the effective potential.
Let us consider a contribution to the effective potential \pref{0,1}
with external lines
$
\psi^{\e_1(\le h)}_{\r_1;x_{0,1},x'_1,l_1}... \psi^{\e_m(\le h)}_{\r_m;x_{0,m},x'_m,l_m}$. By construction the coordinates of the external fields are
such that $(\o x')_{mod 1}\le \g^h$. 
Note that in each graph there is a  
tree of propagators connecting all the vertices and external lines; each propagator carries a coordinate $x$ and vertices connect lines with coordinates differing at most of $\pm 1$; more exactly, if 
$x_i$, $x_j$ are the coordinates of two external lines 
\be x_i-x_j=x'_i+\r_i \bar x_{l_i}-x'_j-\r_j \bar x_{l_j}=
\sum^*_\a \d_\a\label{hhh}\ee 
%; 
where the sum is over the vertices in the path of the tree connecting $i$ and $j$ and $\d_\a=(0,1,-1)$  is associated to the line connected to the vertex $\a$.
When $U=0$ then necessarily  $l_j=l_j$.  It is natural to distinguish among the terms contributing to the effective potential
between {\it resonant terms} and the {\it non resonant terms}. The first are the contributions in \pref{0,1} in which the coordinate $x'$ of the external fields are equal $\psi^{\e_1(\le h)}_{\r_1;x_{0,1},x'_1,l_1}... \psi^{\e_m(\le h)}_{\r_m;x_{0,m},x'_1,l_m}$, that is for any $i.j$
\be
x'_i=x'_j 
\ee
The non resonant terms 
are the ones such that, for some $i,j$, $x'_i\not =x'_j$ so that from \pref{hhh}
and the second Melnikov condition
\pref{d34}
\bea
&&2 \g^{h}\ge ||(\o x'_{i})||+ ||(\o x'_{j})||\ge 
||\o(x'_{i}-x'_{j})||=\\
&&||\o (\r_i\bar x_{l_i}-\r_j\bar x_{l_j})
+\o \sum^*_\a \d_\a ||\ge {C_0\over |\sum^*_\a \d_\a|^{\t'}}\nn
\eea
so that $|\sum^*_\a \d_\a | \ge \tilde C \g^{-h/\t}$.

One can then use the high power or
$J,U$ to get a gain factor making irrelevant the non resonant contributions to the effective potential.
Writing $\e=\max(|J|,|U|)$, $\e=\prod_{h=-\io}^0 \e^{2^{h-1}}
$, we can associate a factor $\e^{2^{h_v-1}}$ for each end-point enclosed in the cluster $v$; as 
$|\sum^*_\a \d_\a |$ is surely smaller that the number of vertices in the cluster $v$ and choosing 
$\g^{1\over\t}/2>1$, we can associate to each non resonant contribution a factor
$ \e^{2^{h-1}|\sum^*_\a \d_\a | }\le \e^{C 2^{h}\g^{-h/\t}}\le \g^{4h}$
for $\e$ small; therefore
\be
\e^{n\over 2}\le \prod_v  \e^{C 2^{h_v}\g^{-h_v/\t}
S^{NR}_v}\le \prod_v \g^{4 h_v S_v^{NR}}
\label{xak}
\ee
where $S_v^{NR}$ is the number of non resonant clusters in $v$; this means that to each non resonant term is associated at least a factor $\g^{4 h_v}$ which is sufficient to make its scaling dimension negative.

It remains to prove that \pref{d33} and \pref{d34} are true in a large relative measure set of values, that is if $|J_\perp|\le \e_0$, in a set of whose complement has
measure $O(C_L \e_0^{1+\a})$, $\a\ge0$,  $C_L$ an $L$-dependent constant.
Indeed if \pref{d33} is true then, if $\cos l\not=0$
\bea
&&C_0 |x|^{-\t}\le ||\o x\pm 2\bar x||
\le ||\o x\pm 2(\bar x+a J_\perp \cos l )||+\nn\\
&&2|a J_\perp \cos l|\le C_0 |x|^{-\t'}+C|\e_0 \cos l a|
\eea
so that  if $\t'>\t+1$, 
$C_0/2 |x|^{-\t}   \le C_0 |x|^{-\t}(1-|x|^{\t-\t'})\le C |a \e_0 \cos l|$ for $|x|\ge 2$ and
and $|x|\ge  (2 C \e_0 \cos l a|/C_0)^{-1\over \t}=\NN_0$. The set $I$ of $J_\perp$ not verifing \pref{d33} is defined by
the condition, for $-1\le s\le 1$
\be f(s)=
\o x\pm 2 (\bar x+J_\perp(s) \cos l a)
= s C_0 |x|^{-\t'}
\label{jj}
\ee
and ${\partial f\over\partial s}={\partial f\over\partial J_\perp}{\partial J_\perp\over\partial s}= C_0 |x|^{-\t'}$
so that the measure of the region in which \pref{d33} is not true is
\bea
&&\int_I dJ_\perp=\sum_l \sum_{n\ge \NN_0} \int_{-1}^1 |{d J_\perp\over ds}| ds
=\nn\\
&&\le \sum_{l, \cos l\not=0} {C\over \cos l}\sum_{x\ge\NN_0}|x|^{-\t'}\le C_ L |\e_0|^{\t'-1\over \t}
\eea
and choosing ${\t'-1\over \t}>1$, that is $\t'>\t+1$ we have that for $|J_\perp|\le \e_0$ the relative measure 
of the excluded $J_\perp $ is $O(C_L \e_0^{\t'-\t-1})$, hence vanishing if $\e_0\to 0$. 
A similar procedure can be repeated for the second Melnikov condition; if $\cos l_i\pm \cos l_j\not=0$ then
\bea
&&C_0 |x|^{-\t}
\le ||\o x\pm (\bar x+a \cos l_i J_\perp )\pm (\bar x+a \cos l_j J_\perp ) ||+
\nn\\
&&|J_\perp a(\cos l_i\pm \cos l_j)|\le C_0 |x|^{-\t'}+C|\e_0 (\cos l_i\pm \cos l_j) a|\nn
\eea
from which
$|x|\ge  (2 C \e_0 |(\cos l_i\pm \cos l_j) a|/C_0)^{-1\over \t}$; one then proceeds as above with 
$|\cos l_i\pm \cos l_j|$ replacing  $|\cos l_i|$.

\section{The resonant terms}

We have seen in the preceding section that the non resonant terms are irrelevant.
We have then to construct a renormalized expansion for the 2-point function, extracting, 
at each RG iteration,
the marginal and relevant part of the resonant terms. In this way
the two-point function is written as an expansion in a set of running coupling constants, 
which is convergent if such constants remain small at each scale; convergence at the end implies localization in the ground state at a non-perturbative level, as it means that the interacting theory is analytically close to the non interacting one, which is localized. 

We focus now on some properties of the resonant terms.
Note that
$x_i-x_j=x'_i-x'_j+\r_i \bar x_ {l_i}-\r_j \bar x_ {l_j}\in\ZZZ$ so that in the resonances
$\r_i \bar x_ {l_i}-\r_j \bar x_ {l_j}\in \ZZZ$. This says that,
up to a zero measure set of $J_\perp$, $\r_i \bar x_ {l_i}-\r_j \bar x_ {l_j}=0$
as 
$(\cos l_i-\cos l_j)a J_\perp$ or $2\bar x+(\cos l_i+\cos l_j) a J_\perp$ 
can be a non vanishing integer only
in a zero measure set
(by the diophantine condition $2\bar x$ cannot be integer). In addition 
in a resonant terms necessarily all the fields have the same $\r$
\be
\r_i=\r_j
\ee
as if $\r_i=-\r_j$ one get $2\bar x+(\cos l_i+\cos l_j)a J_\perp
=0$ which cannot be vanishing for small $J_\perp$. Finally 
if $\cos l_i\not=\cos l_j$
then  necessarily in the resonance $l_j=l_j$, as the condition becomes 
$(\cos l_i-\cos l_j)a J_\perp=0$. %If for instance $L=2$ such condition is always verified. 
The above properties imply that the resonances with a number of fields $\ge 4$ have the following structure
\be
\prod_i \psi^{\e_i}_{\r; x', x_{0,i},l_i }\quad\quad \cos l_j=\cos l_j 
\ee
If $L=2$, that is the array is only composed by two chains
then $l=(0,\pi)$, $\bar x_1=\bar x+J_\perp/2$
and $\bar x_1=\bar x-J_\perp/2$
so that the resonant terms have the same $\r,l$ index; this has the effect that 
the monomials with $\ge 4$ fields and the same coordinates are 
vanishing. In the resonances with a number of fields greater than two there are at least two couples of the form
$\psi^\e_{\r; x',x_{0,1},l}\psi^\e_{\r; x',x_{0,2},l}$ which can be rewritten as
\be
\psi^\e_{\r; x',x_{0,1},l}\psi^\e_{\r; x',x_{0,2},l}=
\psi^\e_{\r; x',x_{0,1},l}(\psi^\e_{\r; x',x_{0,2},l}-\psi^\e_{\r; x',x_{0,1},l})\label{k1}
\ee
and \be
\psi^\e_{\r; x',x_{0,2},l}-\psi^\e_{\r; x',x_{0,1},l}=(x_{0,2}-x_{0,1})\int_0^1 dt \partial \psi^\e_{\r; x',x_{0}(t),l}\label{k2}
\ee
with $x_{0}(t)=x_{0,1}+t(x_{0,2}-x_{0,1})$.
The derivative produces an extra 
$\g^{h_{v'}}$, if $v'$ is the cluster enclosing $v$, and the factor $(x_{0,2}-x_{0,1})$
an extra $\g^{-h_{v}}$; as there are at least two 
of such monomials one gets at least a factor 
$\g^{2(h_{v'}-h_v)}$. Remembering that the scaling dimension is $D=1$, this means that all the resonances  with more than two fields are {\it irrelevant} if $L=2$. 

If $L\ge 3$ the situation is different; there are couple of indices $l,l'$ 
such that $x_l=x_{l'}$; quartic terms involving such couple of indices and the same $x_{0,i}$
are not vanishing so that there are quartic relevant terms. 
For instance in the
three chains problem $L=3$
one has $l=2\pi/3,4\pi/3, 6\pi/3$ and
$\bar x_1=\bar x-J_\perp/2, \bar x_1=\bar x-J_\perp/2, \bar x_3=\bar x$;
the local part of the quartic terms (the part with identical coordinates ) 
$\psi^+_{\r;\xx',1}\psi^-_{\r;\xx',1}\psi^+_{\r;\xx',2}\psi^-_{\r;\xx',2}$
is non vanishing; the quartic terms are indeed relevant while resonant terms with a number greater than $6$ are irrelevant.
The number of couples $i,j$ with $\cos l_i=\cos l_j$, and the corresponding quartic terms, increases with $L$; for instance 
for $L=8$ one
has $l=\pi/4,\pi/2,3\pi/4, \pi, 5\pi/4, 3\pi/2,7\pi/4,2\pi$ with $\cos l=\sqrt{2}/2,0,-\sqrt{2}/2, -1,-\sqrt{2}/2,0,\sqrt{2}/2,1$, so that the non vanishing local quartic terms are $\psi^+_{\r;\xx',1}\psi^-_{\r;\xx',1}\psi^+_{\r;\xx',7}\psi^-_{\r;\xx',7}$, $\psi^+_{\r;\xx',2}\psi^-_{\r;\xx',\r,2}\psi^+_{\r;\xx',6}\psi^-_{\r;\xx',6}$, 
 $\psi^+_{\r;\xx',3}\psi^-_{\r;\xx',3}\psi^+_{\r;\xx',\r,5}\psi^-_{\r;\xx',5}$. As there are at most couples of fields with the same $\bar x_l$ and different $l$, the terms with a number $\ge 6$ of fields are irrelevant, as there are at least $4$ fields with the same $l$.

In order to get a convergent expansion, one has to extract the relevant part from the resonant terms. If $V^h_{res}=\sum_m V_{m,res}^h$ where $V_m^h$ are the monomials with $m$ fields, then we define a localization operation $V^h_2=\LL V_2^h+\RR V^h_2$ with $\RR=1-\LL$ and $\LL$ acts on the kernels of $V_2^h$ in the following way
\be
\LL \hat W^h_2(k_0,x')=\hat W^h_2(0,0)+k_0 \partial_0 \hat W^h_2(0,0)+(\o x')\tilde\partial\ W^h_2(0,0)
\ee
where $\tilde\partial \hat W^h_2(k_0,x')={\tilde W^h_2(k_0,x')-
\tilde W^h_2(k_0,0)\over(\o x')}$. The action of $\RR=1-\LL$ produces a gain $\g^{2(h_{v'}-h_v)}$, using also that 
$(\o x')^2
\sim \g^{2h_{v'}}$, if $v'$ is the smallest cluster enclosing $v$,
for the compact support properties of the lines external to the cluster $v$, while 
$\tilde \partial^2 \hat W^h_2(k_0,x')$ has an extra 
$\g^{-2h_v}$; $\RR V_2^h$ is therefore irrelevant. 
%\be
%\LL \psi^+_{x,x_{0,1},\r,\s}\psi^-_{x,x_{0,2},\r,\s}=
%\psi^+_{x,x_{0,1},\r,\s}T^-_{x,x_{0,1},
%x_{0,2},\r,\s}
%\eehat 
%
%with $T^-_{x,x_{0,1},
%x_{0,2},\r,\s}=
%\psi^{-}_{x',x_{0,1}, \r,\s}+(x_{0,1}-x_{0,2}) 
%\partial\psi^{-}_{x',x_{0,1}, \r,\s}$ 
The local part of the effective potential has then the form
\bea
&&\LL\VV^h_{res,2}=\sum_x {1\over L}\sum_{\r,l} \int d x_0 (\n_{h,l} \g^h\psi^+_{\r;\xx',l}\psi^+_{\r;\xx',l}+\nn\\
&&z_{h,l,\r} \psi^+_{\r;\xx',l}\partial_0 \psi^+_{\r;\xx',l}+\a_{h,l,\r} 
(\o x')\psi^+_{\r;\xx',l} \psi^+_{\r;\xx,l})
\eea
Regarding terms with a number of fields $\ge 6$, $\LL V_{res, m}^h=0$ for $m\ge 6$, as the local part and its first derivative are vanishing.
Finally if $L=1,2$ then
\be
\LL  V_{res, 4}^h=0 
\ee
while for $L\ge 3$ then
\bea
&&\LL V_{res, 4}=G+\\
&&{1\over L^3}\sum_{i, j; \bar x_i=\bar x_j}\sum_\r \l_{h,i,j,\r}\g^h \int d\xx \psi^+_{\r;\xx',i} \psi^-_{\r;\xx',i}
\psi^+_{\r;\xx',j} \psi^-_{\r;\xx',j}\nn\eea
where in $G$ are included marginal terms, that is quartic local terms with at least a field
$\partial \psi$ (the corresponding coupling are called $\tilde\l_{h,i,j,\r}$)
and the sum $\sum_{i, j; \bar x_i=\bar x_j}$ is over the fields with the same $\bar x_l$.

\section{Convergence of the renormalized expansion}

The integration is done separating at each integration step the relevant and the irrelevant part of the effective integration, writing
\be
\int P(d\psi^{\le h})e^{\LL V^{(h)}(\psi^{\le h})+\RR V^{(h)}(\psi^{\le h})}
\ee
with $\RR=1-\LL$ and $\LL$ is the localization operator
defined above; 
this allows to get an expansion in terms of running coupling constants 
$\vec v_h=(\l_{h,l,l',\r}, 
\tilde\l_{h,l,l',\r},\n_{h,l}, \a_{h,l}, z_{h,l})$.
If $v_0$ is the largest cluster, $v$ are the clusters
 (without vertices), 
$\bar v$ the vertices and $R$ or $NR$ the resonant clusters or vertices and $v'$ is the first cluster enclosing $v$, then
$$\prod_{v} \g^{-h_v S_v}=
\prod_{v\not=v_0} 
\g^{-h_{ v'}}\prod
_{\bar v} 
\g^{-h_{\bar v'}}$$
and  
$\prod_{v} \g^{h_v}=\g^{h_{v_0}}
\prod_{v\not= v_0} \g^{h_v}$
so that \pref{paz} can be rewritten as
\be  \e^n\g^{h_{v_0}}
\prod_{v\not=v_0} 
\g^{-(h_{ v'}-h_v)}\prod_{\bar v } 
\g^{-h_{ \bar v'}}\label{xak1}
\ee
Using  \pref{xak} we get
that the kernel
$W^{(h)}_m$ in the renormalized expansion are bounded by, 
if $|v_h|\le \e$ 
\be
\e^{n/2} \prod_v \g^{-h_v(S_v-1)}]\\
[\prod_{v\in R}\g^{2(h_{v'}-h_v)}]
[\prod_{\bar v\in R} \g^{h_{\bar v'}}] [\prod_v \g^{4 S^{NR}_v h_v} ]\label{zaz11}
\ee
where the factor $[\prod_{v\in R}\g^{2(h_{v'}-h_v)}]$ is, as explained in the previous section, due to the action of $\RR$ or to \pref{k1}, \pref{k2}.
Therefore
\pref{zaz11} can be written as
\be C^n \e^{n/2}
 \g^{h_{v_0}}[\prod_{v}\g^{(h_{v'}-h_v)}][\prod_{\bar v\in NR} \g^{h_{\bar v'}}]\label{zaz12}
\ee
As $h_{v'}-h_v\le 0$ it is possible over the scales $h_v$ obtaining a bound from which convergence follows provided that $\e$ is not too large. Note that the above bound is valid for the sum of all Feynman graph of order $n$, by using determinant bounds for fermionic expectations, see  
\cite{M1}, for details. 
The renormalized expansion has a finite radius of convergence 
in terms of the running coupling constants; if they remain, for any $h$, inside the convergence radius then localization is found. We have then to analyze the flow of the effective couplings, and the result is dramatically different in the $L=2$ 
and $L\ge 3$. 

In the case $L=2$ there are no quartic terms in the effective potential, that is $\l_h=\tilde\l_h=0$; the only effective couplings are quadratic and the flow equations are
\be
\n_{h-1,l}=\g\n_h+\b_{h,l}^\n 
\quad \a_{h-1,l}=\a_h+\b_{h,l}^\a 
\ee
and $z_{h-1}=z_h+\b_{h,l}^z$, where $\b_h^\n, \b_h^\a,\b_h^z$ are the beta functions; 
they are given by sum of terms with at least an irrelevant term, as terms containing only
marginal terms (quadratic in the fields) are chain graphs giving a vanishing contribution to be beta function by the compact support properties of the propagator. Therefore, by \pref{zaz12}, 
the beta function is asymptotically vanishing $\b_h^\n, \b_h^\a,\b_h^z=O(\g^h)$.
$\n_{h,l}$ is a relevant coupling but his flow can be controlled by choosing properly the counterterms
$\n_l$; indeed we can write $\n_{h-1,l}=\g^{-h}(\n_l+\sum_{k=h}^0 \g^k \b_{k,l}^\n)$
and choosing $\n_l=-\sum_{k=-\io}^0 \g^k \b_{k,l}^\n$ we get that $\n_{h,l}=O(\g^h\e)$. Moreover 
$\a_{h-1,l}=\sum_{k=h}^0 \b_{h,l}^\a=O(\e)$ and similarly $\a_{h-1,l}=O(\e)$. Therefore if $J,U$ are sufficiently small we have that the running coupling constants are small and the series are convergent.  Similarly if $L\ge 3$   
and $U=0$ there are only quadratic couplings and we can proceed in the same way. 

In the case $L\ge 3$ there are however quartic relevant and marginal couplings, that is 
\be
\l_{h-1,l,l',\r}=\g\l_{h,l,l',\r}+\b_{h,l,l',\r}^\l
\ee
%
%Note that $\l_{0,l,l',\r}=O(U J_\perp)$ as if $J_\perp=0$ there is of course no relevant quartic %terms, as one can easily recognize passing to the coordinate space in the transversal direction. 
Convergence is achieved at finite temperatures, that is for $\g^{-h_\b} U$ or 
$\b U$ of order $1$, and at lower temperatures one expects generically an unbounded flow.

An estimate for the 2-point function follows easily from the expansion for the effective potential; 
if the external coordinates are $x$ and $y$ then there are at least $|x-y|$  $\e$ factors, and this implies exponential decay in the direction of the chains.
By the first Melnikov condition the smallest scale of the contribution ar order $n$ verifies
\be
\g^{-\bar h}\le C (1+\min\{|x|,|y|\})^\t(1+{n\over 1+\min\{|x|,|y|\}})^\t
\ee
from which \pref{zz}  follows.

\section{Conclusions}

We have considered an array of interacting chains with quasi-random disorder. The RG analysis provides an explanation of cold-atoms experiments
\cite{B}, \cite{B1}, in which it is found that localization is present in the single chain case, while is absent when several chains are considered.
In the first case number theoretical properties, combined with cancellations due to Pauli principle, ensure that all the effective interactions are irrelevant, even if dimensionally relevant. On the contrary,
in the second case there are non vanishing relevant interactions, whose number increases
with the number of chains; as usual, the presence of diverging directions in the RG flow is expected to signal an instability of the system. 
In addition, we have shown that localization in the ground state is present with two chains, if the fermions are spinless and in presence of interaction, a prediction
in principle accessible at an experimental verification.

\end{document}